\documentclass[12pt]{article}
\usepackage{graphicx}
\usepackage{amsmath,amssymb}
\usepackage[
dvipdfm,truedimen,
top=25truemm,bottom=20truemm,%
left=20truemm,right=15truemm]{geometry}
\usepackage{multicol}
\usepackage{physics}
\usepackage{cite}
\usepackage{titlesec}
\numberwithin{equation}{section}
\usepackage{mathrsfs}
\usepackage{tikz}
\usepackage{enumitem}
\usepackage{subcaption}
\usepackage{hyperref}
\usepackage[percent]{overpic}

\begin{document}

\title{Page curves and island's delays in asymptotically flat 2d spacetimes with injections}

\author{Yuuta Saito\\
{\small Graduate School of Science and Technology, Nihon University,} \\{\small 1-8-14, Kanda-Surugadai, Chiyoda-ku Tokyo 101-8308, Japan}\\ \texttt{\small csyt23002@g.nihon-u.ac.jp}
}
\date{}

\maketitle
\begin{abstract}%
We explore spacetimes with multiple energy injections in asymptotically flat two-dimensional black hole and analyze the entanglement entropy in such spacetimes.
This work is an extension of the setup of the single-injection case, by F.~F.~Gautason, L.~Schneiderbauer, W.~Sybesma and L.~Thorlacius, to include the multiple energy injections. We derive the solution of the model, by J.~G.~Russo, L.~Susskind and L.~Thorlacius, for a general number $n$ of the total injections, and discuss the entropy only for the case $n=2$. 
The essential point of this work is in the delay of the island. This delay makes the intermediate state necessary, where the island's endpoint lies between the 1st and the 2nd injections while the observer is located after the 2nd injection. The intermediate state makes the entanglement entropy evolve continuously across the 2nd injection time.
\end{abstract}

\section{Introduction}
A black hole is a region from which nothing can escape. In general relativity, it is described by Einstein's equation, or more precisely, by the metric equations of motion derived from the action. However, when quantum effects are taken into account, black holes emit thermal radiation, known as the Hawking radiation \cite{Hawking:1975vcx}. As a result of this radiation, a black hole evaporates by emitting its energy into infinity. Thus, the number of states of the black hole is expected to decrease, since it is believed to be proportional to exponential of the area of the event horizon. 

On the other hand, the quantum entanglement between the inside and the outside of the black hole increases due to the pair creation near the horizon. The strength of this entanglement is quantified by the entanglement entropy, defined as the von Neumann entropy of the reduced density matrix. Moreover, this quantity is bounded above by the logarithm of the smaller number of states between the black hole and the radiation. 

Intuitively, the entanglement entropy of the Hawking radiation increases monotonically. However, the upper bound starts to decrease once the number of states in the black hole interior and in the outside region become equal. Therefore, we wonder whether the monotonically increasing entanglement entropy exceeds its upper bound. In a unitary theory, such an inconsistency must not occur. If the entropy exceeds the bound, this contradiction is called the information paradox \cite{Hawking:1976ra}. If the paradox doesn't occur, the entropy should follow the Page curve \cite{Page:1993wv,Page:2013dx}, which describes its time evolution: it increases initially and then decreases after a specific time, known as the Page time. 

Recently, a method to resolve the paradox has been proposed —  the so-called island formula \cite{Almheiri:2019hni}.
 It provides a prescription for computing the entanglement entropy in gravitational systems, given by
\begin{equation}
S_\text{\bf rad}({\rm R})=\underset{\mathscr{I}}{\text{min}}\cdot\underset{\mathscr{I}}{\text{ext}}\left[{{\rm Area}(\partial \mathscr{I})\over{4G_N}}+S_{\rm ent}({\rm R}\cup \mathscr{I})\right]=\underset{\mathscr{I}}{\text{min}}\cdot\underset{\mathscr{I}}{\text{ext}}\,S_\text{gen}({\rm R},\mathscr{I}). \label{island}
\end{equation} 
Here, $S_\text{\bf rad}({\rm R})$ denotes the entanglement entropy of the region R in quantum gravity. The region $\mathscr{I}$ is called the island, and $S_{\rm ent}({\rm R}\cup \mathscr{I})$  is the entanglement entropy of the union of R and $\mathscr{I}$, computed using quantum field theory on a fixed background. $G_N$ is the Newton's constant. The symbol $\underset{\mathscr{I}}{\text{min}}\cdot\underset{\mathscr{I}}{\text{ext}}$ means that the quantity inside the brackets is to be extremized with respect to $\mathscr{I}$, and the minimum value among the extrema is taken. 
The second term in the brackets plays the main role in canceling the quantum entanglement, since the quantum partners are observed in both regions R and $\mathscr{I}$. The first term, on the other hand, behaves similarly to the decreasing upper bound on the entanglement entropy. 
If there is no island, the first term vanishes, and the second term is evaluated only in the region R.
Moreover, the ``min" in the island formula indicates that, at each moment, the entanglement entropy is given by the smaller of the two: the value without the island and the one with the island. Roughly speaking, the entanglement entropy is given by min$[S_{\rm gen}({\rm R},\varnothing),\,{\text{ext}_{\mathscr{I}}}S_\text{gen}({\rm R},\,{\mathscr{I}})]$. 
Thus, in the absence of an island, the entanglement entropy increases, whereas in the presence of an island, it decreases.
Through this mechanism, the island formula reproduces Page curve. In fact, several papers have shown that island formula leads to the Page curve \cite{Almheiri:2019hni,Almheiri:2019yqk,Hashimoto:2020cas,Goto:2020wnk,Almheiri:2019psy,Wang:2021woy,Gautason:2020tmk,Anegawa:2020ezn, Hartman:2020swn,RoyChowdhury:2022awr}, thereby resolving the information paradox. 

In the previous study \cite{Gautason:2020tmk}, the Russo--Susskind--Thorlacius (RST) model \cite{Russo:1992ax} has been used to analyze the island formula. This model describes a two-dimensional dilaton gravity theory in asymptotically flat spacetimes and incorporates the backreaction of quantum effects into the classical Callan-Giddings-Harvey-Strominger (CGHS) model \cite{Callan:1992rs}. In \cite{Gautason:2020tmk}, the initial spacetime is flat. By the energy injection from the infinity, gravitational collapse occurs, resulting in the formation of a black hole. This energy injection makes the entanglement entropy increase between the black hole region B and the radiation region R. In the following, we review this scenario and demonstrate how the island formula shows the Page curve and Page time. However, the paper \cite{Gautason:2020tmk} computes the entanglement entropy not for the radiation region but for the black hole region B.
Since we typically assume a pure state when discussing the information paradox, it does not matter whether we compute the entanglement entropy in the black hole region or the radiation region, in order to derive the Page curve. 

In this paper, we extend the setup of \cite{Gautason:2020tmk} to include multiple energy injections.\footnote{This work is also motivated by the suggestion that a black hole might not form when the multiple shells shrink in four-dimensional spacetime \cite{Kawai:2013mda}, thereby avoiding the information paradox \cite{Kawai:2015uya}. In our two-dimensional model, however, the horizon does form and even moves due to the injection. The reason for this is not clear, but it may be related to the fact that our model is two-dimensional.} The original analysis assumes a single injection, but black holes in realistic scenarios may experience repeated or prolonged matter influx. To better understand the behavior of the island and the entanglement entropy under such conditions, we analyze how they respond to sequences of injections. As in the original setup, we place the trajectory of an observer as the anchor curve, which indicates the boundary between the black hole region and the radiation region. Although each behavior of the entanglement entropy at early and late time (after all injections) respectively matches that of the single-injection case, it differs in the intermediate phase. More precisely, we confirm that the endpoint of the island goes beyond the trajectory of the injection later than the anchor curve, namely, the time of the island delays compared with the one of the anchor curve. 
This delay guarantees the entanglement entropy to evolve continuously across each injection time.
Interestingly, after each injection, the entanglement entropy increases again, and eventually decreases once more when the endpoint of the island finally moves from the pre-injection spacetime to the post-injection spacetime.

We organize this paper as follows. In the remainder of section 1, we briefly review the setup and the Page curve in the case of a single injection \cite{Gautason:2020tmk}. In section 2, we construct solutions with multiple energy injections in the RST model. In section 3, we apply the island formula to the solution in section 2, and derive the Page curve analytically using approximations. In this calculation, we consider the two injections in total. In particular, we consider the two cases. First, the first appearance of the island's endpoint is located after the 2nd injection. Second, the 2nd injection is after the Page time. In addition, we compute the Page curve numerically. Finally, in section 4, we summarize our results and discuss the future work.

\subsection{Review of the Page curve in the single-injection case \cite{Gautason:2020tmk}}\label{single}
In the remainder of section 1, we consider the RST model in asymptotically flat 2-dimensional black hole spacetimes. In this model, the dilaton field, which is coupled to gravity, effectively plays the role of the metric. Therefore, we understand the spacetime geometry by solving the equation for the dilaton. We consider the flat spacetime before a single injection, and the black hole spacetimes after it. Thus, the metric takes the following form:
\begin{align}
ds^2_{\rm flat}&={du\,dv\over{uv}}\hspace{2,3cm}(v<1),\label{flat}\\
         ds^2_{\rm bh}&=-{du\,dv\over{1-v(1+u)}}\hspace{1cm}(v\ge 1).
\end{align} 
Here, $u$ and $v$ are the ingoing and outgoing null coordinates, respectively. $v=1$ corresponds to the trajectory of the injection.

In this spacetime, the entanglement entropy without the island, calculated in the conventional way, is given by 
\begin{equation}
S_\text{gen}^\text{no-island}={c\over{12}}\log\left|{\log\left({u_{\rm A}\over{u_{\rm bdy}}}\right)^2\log\left({v_{\rm A}\over{v_{\rm bdy}}}\right)^2{u_{\rm A} v_{\rm A}\over{1-v_{\rm A}(1+u_{\rm A})}}}\right|. \label{def}
\end{equation} 
Here, $c$ is the central charge, and  we adopt the notation $S_\text{gen}^\text{no-island}$ by following \cite{Gautason:2020tmk}. In the above discussion, it is expressed as $S_{\rm gen}({\rm R},\varnothing)$.
In \eqref{def}, the entropy is calculated by $S_{\rm ent}({\rm B})$, using the relation $S_{\rm gen}({\rm R},\varnothing)=S_{\rm ent}({\rm R})=S_{\rm ent}({\rm B})$.  Each of the subscripts ``A'' and ``bdy" refers to the point on the anchor curve and on the spacetime boundary, respectively. On the other hand, we introduce the coordinates $(t,\, \sigma)$, which are defined  by $u=-1-{\rm e}^{-t+\sigma}$ and $v={\rm e}^{t+\sigma}$.
Substituting these coordinates into the metric, we find $ds^2\sim dt^2-d\sigma^2$ at $\sigma\sim\infty$, and the coordinates cover the outside of the horizon. We define the anchor curve as the line with a constant value $\sigma=\sigma_{\rm A}$, while $t$ is still a variable corresponding to a time coordinate on the anchor curve, and we call it $t_{\rm A}$ in the following discussion. Then we define $(u_{\rm A} ,v_{\rm A})=(-1-{\rm e}^{-t_{\rm A}+\sigma_{\rm A}},\,{\rm e}^{t_{\rm A}+\sigma_{\rm A}}$). We assume that the injection energy $M$ is much larger than the central charge, i.e. $\epsilon ={c/{(48 M)}}\ll 1$. 
As we will see shortly, if there exists the island, we must consider the regime e$^{t_{\rm A}-\sigma_{\rm A}}\gtrsim1/\epsilon$.
Then, \eqref{def} is approximately expressed as  
\begin{equation}
S_\text{gen}^\text{no-island}=4M\epsilon(t_{\rm A}-\sigma_{\rm A})+\cdots.\label{single no island}
\end{equation} 
Focusing on the leading contribution, it grows linearly in time.

\begin{figure}[htbp]
  \centering
  \begin{overpic}[width=0.6\linewidth]{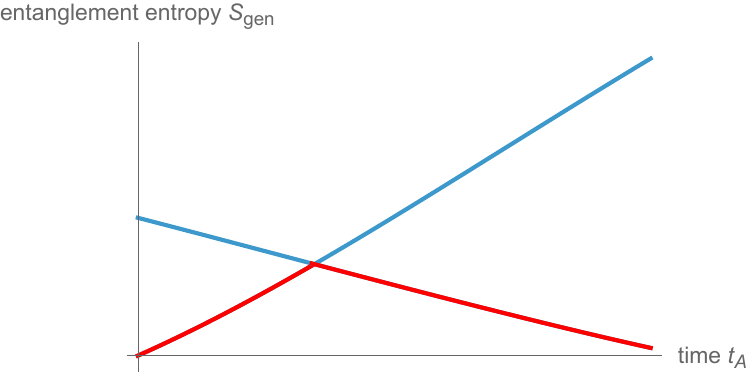}
    \put(41.8,3){\dashline{1}(0,0)(0,31)} 
    \put(33,35){Page time}
    \put(58,36){no island}
    \put(58,12){with island}
  \end{overpic}
  \caption{The Page curve for the single-injection case,\cite{Gautason:2020tmk}. 
  Here, in the numerical computation, we use the following parameters: $M=3$, $\epsilon = 1/15$, $x_1 =1$,  $\sigma_{\rm A}=2$. 
  The increasing and decreasing line correspond to the cases without and with the island respectively. 
  The red line expresses the minimum value at each time, so is the Page curve. }
  \label{1p}
\end{figure}
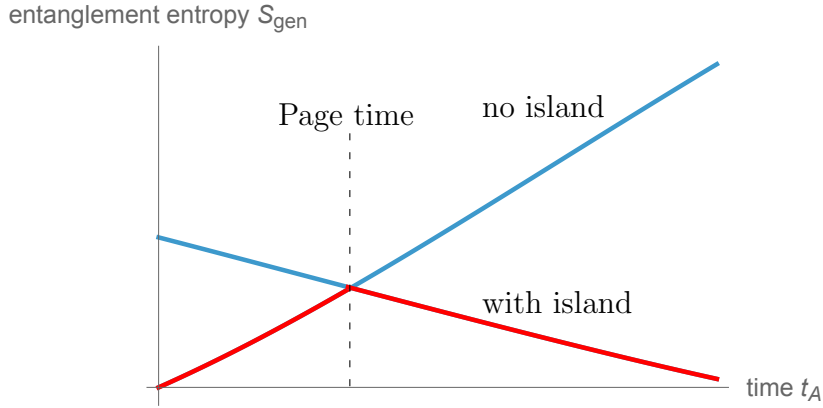

On the other hand, considering the case with the island, the generalized entropy is given by
\begin{align}
S_\text{gen}^{\rm island}=&2M\{1-v_{\rm I }(1+u_{\rm I })-\epsilon\log(-Mv_{\rm I }u_{\rm I })\}\notag\\
+&{c\over{12}}\log\left|{\log\left({u_{\rm A}\over{u_{\rm I }}}\right)^2\log\left({v_{\rm A}\over{v_{\rm I }}}\right)^2{u_{\rm A} v_{\rm A}\over{1-v_{\rm A}(1+u_{\rm A})}}{u_{\rm I } v_{\rm I }\over{1-v_{\rm I }(1+u_{\rm I })}}}\right|. \label{1e}
\end{align} 
Here, the left-hand side represents $S_\text{gen}({\rm R},\,{\mathscr{I}})$, while the matter term in \eqref{island} is calculated by using $S_{\rm ent}({\rm B})$. The subscript ``\,I\,'' denotes the endpoint I on the island. 
The first term corresponds to the area term in the island formula \eqref{island}, for which we adopt the Wald entropy \cite{Wald:1993nt,Myers:1994sg}.
According to the island formula, we determine the position of the island's endpoint I by extremizing this expression with respect to $(u_{\rm I },\,v_{\rm I })$. Then, we obtain that the endpoint I of the island is located at 
$u_{\rm I} +1 \sim {\rm e}^{-t_{\rm A}+\sigma_{\rm A}}$,
$v_{\rm I} \sim  3\epsilon{\rm e}^{t_{\rm A} - \sigma_{\rm A}}$.
We assume that the island appears with its endpoint located after the energy injection, i.e. $v_{\rm I}>1$, so we must be in the regime e$^{t_{\rm A}-\sigma_{\rm A}}\gtrsim1/\epsilon$.  
As a result, we can express the entropy as 
\begin{equation}
S_\text{gen}^{\rm island}=2M-2M \epsilon(t_{\rm A}-\sigma_{\rm A})+\cdots,\label{singleinjectionisland}
\end{equation}
which decreases monotonically in time. By comparing this with the no-island behavior \eqref{single no island}, the Page  time is obtained as 
\begin{equation}
t_p:=(t_{\rm A}-\sigma_{\rm A}){\Big |}_{S_\text{gen}^\text{no-island}=S^{\rm island}_\text{gen}}={1\over{3\epsilon}}.\label{tpagesingle}
\end{equation}
 Moreover, by the numerical computation, the Page curve is derived \cite{Gautason:2020tmk}, as shown in figure \ref{1p}.
In the following sections, we carry out the parallel analysis for the two-injection case.

\section{RST model with multiple injections}
\subsection{RST model}
The CGHS model \cite{Callan:1992rs} is one of the most important works in two dimensional gravity. The action is given by
\begin{equation}
I_{\rm CGHS}={1\over{2\pi}}\int d^2x \sqrt{-g}\, {\rm e}^{-2\phi}[R+4(\nabla \phi)^2+4\lambda^2].
\end{equation}
Here, $g$ is the determinant of the metric $g_{\mu \nu}$, $R$ is the scalar curvature, $\lambda^2$ is the cosmological constant and $\phi$ is the dilaton, which is the scalar field coupled to the scalar curvature.
In this paper, $\lambda$ is set to 1. 
In addition, we consider the conformal field theory (CFT) coupled to the gravity. Varying the metric or the dilaton, we derive the equations of motion:%
\footnote{
We define the energy momentum tensor as follows:
\begin{equation}
T_{\mu\nu}=-{4\pi\over{\sqrt{-g}}}{\delta S\over{\delta g^{\mu\nu}}}.\notag
\end{equation}
This definition is different from CGHS paper \cite{Callan:1992rs} by a factor of $-2$. 
}
\begin{align}
4{\rm e}^{-2\phi} [\nabla_\mu\nabla_\nu \phi - g_{\mu\nu}(\nabla^2 \phi-(\nabla \phi)^2+1 )]&=T_{\mu\nu}^{\rm cl},\\
R-4(\nabla\phi)^2+4\nabla^2\phi+4&=0.\label{p}
\end{align} 
For simplicity, we consider the conformal gauge:
\begin{equation}
ds^2=-{\rm e}^{2\rho}dx^+dx^-.
\end{equation}
Using this metric, we rewrite the action and the equations of motion as follows:
\begin{align}
I_{\rm CGHS}={1\over{\pi}}\int dx^2 &\left(2{\rm e}^{-2\phi}\partial_+\partial_-\rho-4{\rm e}^{-2\phi}\partial_+\phi\partial_-\phi +{\rm e}^{2(\rho-\phi)}\lambda^2\right), \label{rs}\\
&\partial_{+}\partial_{-}{\rm e}^{-2\phi}+{\rm e}^{2(\rho-\phi)}=0,\label{+-class}\\
-2\partial_{\pm}\partial_{\pm}{\rm e}^{-2\phi}&+4(\partial_{\pm}{\rm e}^{-2\phi})(\partial_{\pm}\rho-\partial_{\pm}\phi)=T_{\pm\pm}^{\rm cl},\label{++--class} \\
2{\rm e}^{-2\rho}\partial_{+}\partial_{-}(\rho-\phi)&+2{\rm e}^{-2\rho}(-\partial_{+}\partial_{-}\phi+2\partial_{+}\phi\partial_{-}\phi)+1=0\label{pi}
.
\end{align}
Here, \eqref{rs} possesses a symmetry under the transformation $\rho\rightarrow\rho+\epsilon e^{2\phi},\,\phi\rightarrow\phi+\epsilon e^{2\phi}$, which leads to the current conservation equation, $\partial_+\partial_-(\rho-\phi)=0$. This conservation is also derived from \eqref{+-class} and \eqref{pi}. 
This conservation equation allows the conformal factor $\rho$ to be identified with the dilaton $\phi$, which makes \eqref{+-class} and \eqref{++--class} simple. Therefore, the classical geometry can be derived by obtaining the solution $\phi$. 

Next, we consider the quantum effects arising from the conformal anomaly,
 \begin{equation}
\langle T^\mu_\mu\rangle={c\over{12}}R \implies  \langle T_{+ -}\rangle=-{c\over{6}}\partial_+\partial_-\rho,\label{pm}
\end{equation}
which induces the backreaction on the geometry. Using this anomaly and the conservation law $\nabla_\mu T^{\mu}_{\nu}=0$, the other components are derived:
 \begin{equation}
 \langle T_{\pm \pm}\rangle={c\over{12}}\left\{2\partial_{\pm}^2\rho-2(\partial_{\pm}\rho)^2+t_{\pm}\right\}\label{tpm}.
\end{equation}
Here, $t_{\pm}$ are functions of integration depending on $x^\pm$, which reflect quantities related to boundary conditions of quantum matter. Under the coordinate transformation $x^{\pm} \rightarrow y^{\pm}(x^{\pm})$, the conformal factor $\rho(x^+,x^-)$ transforms as $\tilde{\rho}(y^+,y^-)=\rho(x^+,x^-)-{1\over{2}}\log{{dy^+\over{dx^+}}{dy^-\over{dx^-}}}$.
Due to this transformation property, $t_{\pm}$ transform as
\begin{equation}
\left({dy^{\pm}\over{dx^{\pm}}}\right)^2\tilde{t}_{\pm}(y^{\pm}) =t_{\pm}(x^{\pm})+\{y^{\pm}(x^{\pm}),x^{\pm}\}.\label{t-t}
\end{equation}
Here, the second term on the right-hand side is the Schwarzian derivative:
\begin{equation}
\{y(x),x\}={y'''(x)\over{y'(x)}}-{3\over{2}}\left({y''(x)\over{y'(x)}}\right)^2.\label{sch}
\end{equation}
According to \cite{Callan:1992rs}, the energy currents \eqref{tpm} can be interpreted as thermal radiation, that is, Hawking radiation in the future null infinity.
These quantum effects \eqref{pm}, \eqref{tpm} can be captured by adding the non-local Polyakov term to the action:
\begin{align}
I_\text{Polyakov}&=-{c\over{96\pi}}\int d^2x \sqrt{-g}R{1\over{\nabla^2}}R,\label{np}\\
                           &={c\over{12\pi}}\int d^2x \partial_{+}\rho\partial_{-}\rho.\label{ppm}
\end{align}
Here, we take the conformal gauge at the second equality.
However, this term breaks the symmetry of the action mentioned above. 
To recover the symmetry, the paper \cite{Russo:1992ax} introduces the following term,
\begin{equation}
 I_\text{RST}=-{c\over{48\pi}}\int d^2x\sqrt{-g}\phi R.
\end{equation}
As a result, the action $I= I_\text{CGHS}+I_\text{Polyakov}+ I_\text{RST}$ possesses the symmetry that the conformal factor $\rho$ identified with the dilaton $\phi$, while the transformation is different from the above one, $\rho\rightarrow\rho+\epsilon {\rm e}^{2\phi}/(1-{c\over{48}}{\rm e}^{2\phi}),\,\phi\rightarrow\phi+\epsilon {\rm e}^{2\phi}{/(1-{c\over{48}}{\rm e}^{2\phi})}$. Thus, we let the conformal factor be regarded as the dilaton. From the action with the quantum effects, we get the equation of motion. In order to solve this equation easily, we introduce the quantity,
\begin{equation}
\Omega={\rm e}^{-2\phi}+{c\over{24}}\phi.\label{od}
\end{equation}
Using this quantity, we get the equations as follows, 
\begin{align}
\partial_+\partial_- \Omega&=-1,\label{O1}\\
-2\partial_{\pm}^2 \Omega&={c\over{12}}t_{\pm}+T^{\rm cl}_{\pm\pm}.\label{O2}
\end{align}
Here, 
$t_{\pm}$ are the quantities appeared in \eqref{tpm}. In the case without the flux at the past null infinity except for the energy injection, $t_{\pm}$ can be written as follows:
 \begin{equation}
t_{\pm}=-{1\over{2(x^{\pm}})^2}.\label{ttpm}
\end{equation}
Under the argument like \cite{Gautason:2020tmk}, we consider the energy momentum tensor as 
\begin{equation}
T^{\rm cl}_{++}=\sum_{k=1}^{n}{2E_{k}\over{x_{k}}}\delta(x^+-x_k), \hspace{1cm} T^{\rm cl}_{--}=0.\label{TT}
\end{equation}
In the $x^{\pm}$ coordinates, the above energy momentum tensors mean that there are no outgoing currents, and that the ingoing energy injection $E_{k}$ is at $x^{+}=x_k$.  
Using \eqref{ttpm} and \eqref{TT}, we derive the solution $\Omega$ from \eqref{O1} and \eqref{O2},
\begin{equation}
\Omega=-x^+x^--{c\over{48}}\log(-x^+x^-)-\sum_{k=1}^{n}{E_{k}\over{x_k}}(x^+-x_k)\Theta(x^+-x_k).
\end{equation}
Here, as for the initial state, we assume that the spacetime before the first injection is flat, which is described by the solution so-called the linear dilaton,
\begin{equation}
\Omega=-x^+x^--{c\over{48}}\log(-x^+x^-). \label{ino}
\end{equation}
From the definition of $\Omega$, \eqref{od}, this initial state enables us to derive the dilaton e$^{-2\phi}=-x^+x^-$, and the spacetime is flat as mentioned at \eqref{flat}.

\subsection{Properties of the solution with multiple injections}
\subsubsection{The apparent horizon of the solution with multiple injections}
The horizon of a black hole can be defined in several ways. One such definition is the apparent horizon.
In a mathematically rigorous definition, the apparent horizon is defined as the boundary of a trapped region, where the expansion of outgoing future-directed null geodesic congruence vanishes. The trapped region itself is characterized by the expansion of outgoing future-directed null geodesic congruence being negative.
In dilaton gravity, this structure can alternatively be described by using the dilaton $\phi$ \cite{Russo:1992ht}.
 The apparent horizon corresponds to the surface on which the covariant derivative $\nabla_\mu \phi$ becomes null, 
and the region where $\nabla_\mu \phi$ is timelike is identified with the trapped region.

In this framework, the apparent horizon is given by the curve
\begin{equation}
x^-=-{c\over{48}}{1\over{x^+}}-\sum_{k=1}^n{E_k\over{x_k}}\Theta(x^+-x_k).\label{aph}
\end{equation} 
We see that $x^-$ discontinuously decreases due to each injection. 
Since we consider the region $x^-<0$, this behavior represents the expansion of the horizon by the each injection.

\subsubsection{A semiclassical limit of the solution with multiple injections}
We will consider a semiclassical limit. 
In this limit, the injection energy is much larger than the central charge. Thus, we define the approximation parameter as $\epsilon\mathrel{:=}c/(48 E_{1})\ll 1$, following the convention in \cite{Gautason:2020tmk}. Moreover, we assume that the injected energies are of the same order. Therefore, 
we have $\epsilon\sim c/(48 E_{2})\sim \cdots \sim c/(48 E_{n})$, where $n$ is the number of injections. 
Using this limit, the leading part of $\Omega$ is expressed as 
\begin{equation}
\Omega=E_1\left[\sum_{k=1}^{n}{E_{k}\over{E_1}}\Theta(x^+-x_k)-v\left(\sum_{k=1}^{n}{E_{k}\over{E_1}}{x_{1}\over{x_k}}\Theta(x^+-x_k)+u\right)\right]+\order{\epsilon}.
\end{equation}
Here, we have introduced the coordinate
\begin{equation}
x^+=x_1 v,\hspace{1cm}x^-={E_1\over{x_1}}u,\label{xxuv}
\end{equation}
again following  \cite{Gautason:2020tmk}. In the coordinate, $v=x_k/x_1$ corresponds to the time of the $k$th energy injection. In particular,  $v=1$ corresponds to the time of the first injection.
Comparing this expression with \eqref{od}, the metric becomes 
\begin{equation}
ds^2=-{dudv\over{\sum_{k=1}^{n}{E_{k}\over{E_1}}\Theta(x^+-x_k)-v\left(\sum_{k=1}^{n}{E_{k}\over{E_1}}{x_{1}\over{x_k}}\Theta(x^+-x_k)+u\right)}}+\order{\epsilon}.\label{semimetric}
\end{equation}
From the conventional form of the 2d metric \cite{Mandal:1991tz,Witten:1991yr}, we find that the event horizon lies at the position  
$\sum_{k=1}^{n}{E_{k}\over{E_1}}{x_{1}\over{x_k}}+u=0$, which coincides with the apparent horizon after $n$th injection in this limit. On the other hand, the spacetime singularity is located at the position that the denominator of \eqref{semimetric} vanishes.
Later, we will use this metric to compute the entanglement entropy via the island formula.

\section{Page curves}
In the vacuum state of the CFT, the entanglement entropy can be computed by using the replica trick or holography.
We assume that the CFT is in the vacuum state when the metric is $ds^{2}={\rm e}^{2\Phi}dy^+dy^-$.
In the RST model, the coordinates $y^{\pm}$ are specified by the condition $\tilde{t}_{\pm}(y^{\pm})=0$. Then, we have the coordinate transformation $v={\rm e}^{y^+},\,u=-{\rm e}^{-y^-}$ by \eqref{t-t}, \eqref{ttpm} and \eqref{xxuv}.
Under this transformation, the conformal factor in the vacuum state ${\rm e}^{2\Phi}$ is given by
 \begin{equation}
{\rm e}^{2\Phi}={vu\over{\sum_{k=1}^{n}{E_{k}\over{E_1}}\Theta(x^+-x_k)-v\left(\sum_{k=1}^{n}{E_{k}\over{E_1}}{x_{1}\over{x_k}}\Theta(x^+-x_k)+u\right)}}+\order{\epsilon}. \label{epi}
\end{equation}
If we consider the reduced density matrix in the spacelike interval D$\mathrel{:=}[p,q]$, 
then the matter contribution to the entanglement entropy is given by\footnote{The paper \cite{Gautason:2020tmk} derives this formula using a formula from the holography \cite{Ryu:2006bv,Hubeny:2007xt}, which itself is derived from the AdS/CFT correspondence \cite{Maldacena:1997re}. Another way to derive this formula is the replica method \cite{Calabrese:2004eu}.} 
\begin{equation}
S_\text{ent}({\rm D})={c\over{12}}\log\left|{l^4({\rm D}) {\rm e}^{2\Phi(p)}{\rm e}^{2\Phi(q)}}\right|.\label{Sm}
\end{equation}
Here, $l({\rm D})$ denotes the proper distance between $p$ and $q$. 
Later, we will use the entanglement entropy \eqref{Sm} with the conformal factor \eqref{epi} to analyze the Page curve via the island formula. 
Specifically, we consider $n=2$.
We use the anchor curve which is different from \cite{Gautason:2020tmk} before the 2nd injection. We assume that the difference does not affect much the essential behavior of the entropy.
Then we mainly focus on the case in which the point A on the anchor curve  is located after the $2$nd energy injection, i.e., $v_{\rm A}\ge {x_2/x_1}$, and study the entropy by following \cite{Gautason:2020tmk}. 

Then we compute three cases: without the island, with the island's endpoint I located after the 2nd injection, and with the island's endpoint I located between the 1st and the 2nd injections. Hereafter, we refer to the third case as the intermediate state. 
 
\begin{figure}[htbp]
  \centering
  \includegraphics[width=0.292\linewidth]{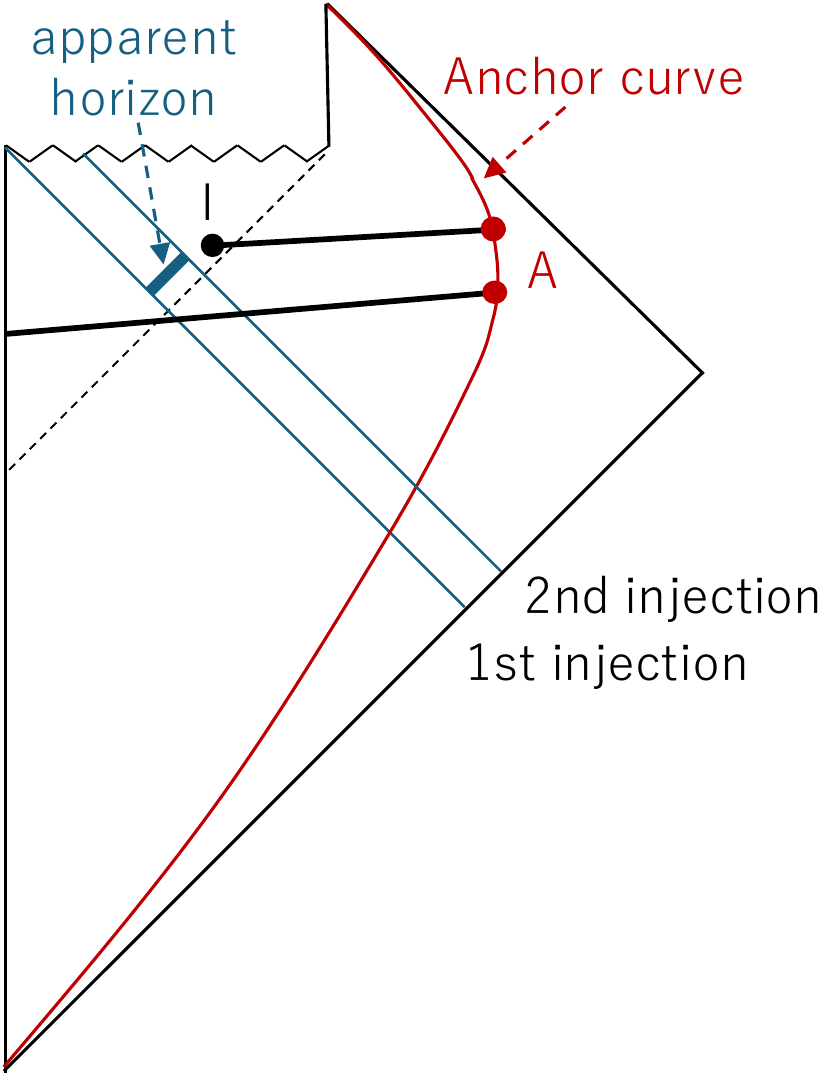}
  \caption{The Penrose diagram illustrating a two-injection setup. The red line indicates the anchor curve. The blue thin lines correspond to the energy injections. The dashed line represents the event horizon, and the blue thick line denotes the apparent horizon. This case corresponds to the short period between the 1st and the 2nd injections, which means that the first appearance of the island's endpoint is located after the 2nd injection. Therefore, after the 2nd injection, the case without the island is initially favored, and then the case with the island is selected after the Page time. The situation is similar to the single-injection case, in that after all injections we only need to consider two possibilities: the no-island case and the case with the island.}
  \label{pd1}
\end{figure}

\begin{figure}[htbp]
  \centering
  \includegraphics[width=0.292\linewidth]{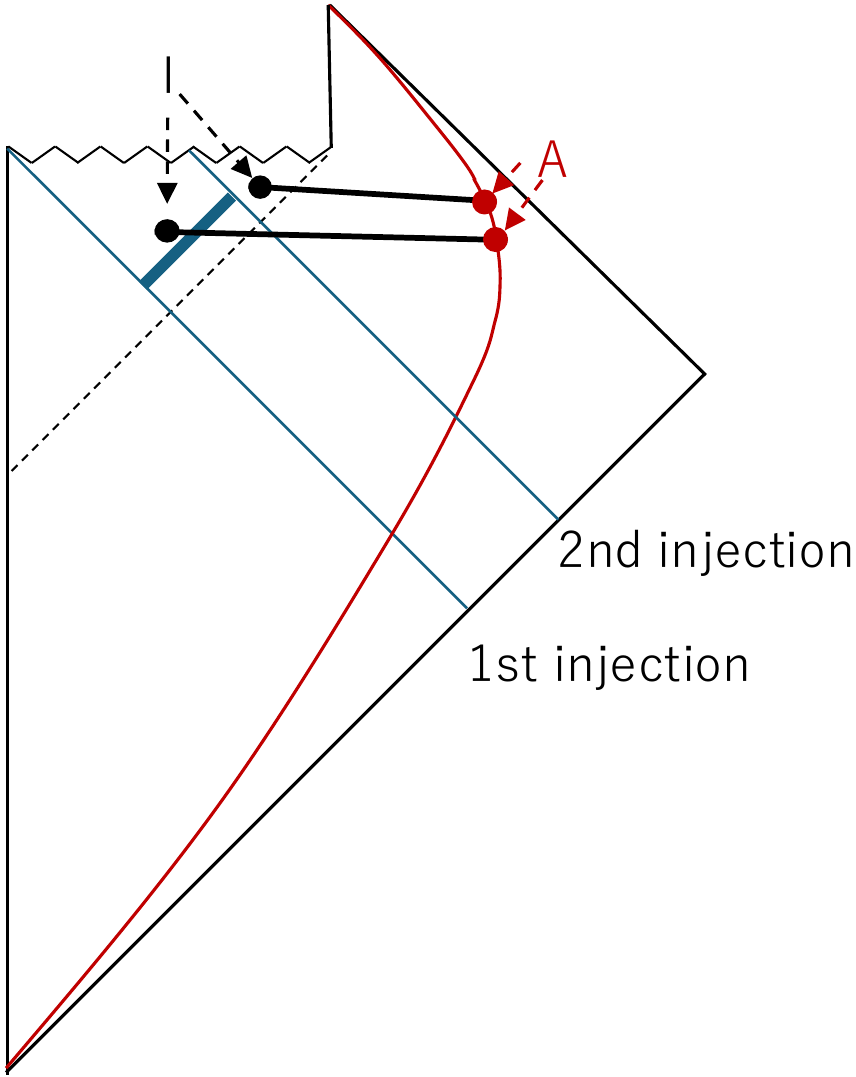}
  \caption{The Penrose diagram illustrating another two-injection setup. The interpretation of the diagram is the same as in figure~\ref{pd1}.
In this case, the 2nd injection occurs after the island appears. Therefore, we consider 2 cases; (i) the island's endpoint exists in the region between the 1st injection and the 2nd injection— this will be referred to as the intermediate state in the main text — and (ii) in the region after the 2nd injection.}
  \label{pd2}
\end{figure}

For the discussion of the Page time, we consider two cases. The first corresponds to the short period between the 1st and the 2nd injections, as shown in figure~\ref{pd1}. More precisely, the 2nd injection occurs before the Page time, and the first appearance of the island's endpoint is after the 2nd injection. The second is the case where the 2nd injection occurs after the Page time. Figure~\ref{pd2} shows the situations some time after the Page time in the second case, when the point A on the anchor curve has moved to the region after the 2nd injection.  
These arguments can be generalized to two cases when we consider totally $n$ injections. 
The first is the case that the first appearance of the island's endpoint is after the $n$th injection.
The second is the case where the first appearance of the endpoint is between the $(n-1)$th and the $n$th injection, while the point A on the
anchor curve is before the nth injection at that moment.

\subsection{Without island} 
When computing the entanglement entropy without the island, we use the expression \eqref{Sm} only.
Accordingly, we obtain 
\begin{equation}
S_{\rm gen}^\text{no-island}={c\over{12}}{\log}\left|\left(\log{v_{\rm A}\over{v_{\rm bdy}}}\log{u_{\rm A}\over{u_{\rm bdy}}}\right)^2 {\rm e}^{2\Phi({\rm bdy})}{\rm e}^{2\Phi({\rm A})}
\right|.\label{ni2i}
\end{equation}
As mentioned earlier, the subscripts “bdy” and “A” refer to the coordinates of the spacetime boundary satisfying $v_{\rm bdy} = -\epsilon/u_{\rm bdy}$, and a point at the anchor curve, respectively.
The coordinates $v_{\rm A}$ and $u_{\rm A}$ take the following form: 
\begin{align}
v_{\rm A}&=v_{\rm A}(t_{\rm A})={\rm e}^{t_{\rm A}+\sigma_{\rm A}},\label{uva1}\\
u_{\rm A}&=u_{\rm A}(t_{\rm A})=-1-{E_2\over{E_1}}{x_1\over{x_2}}-{\rm e}^{-t_{\rm A}+\sigma_{\rm A}}=-1-{E_2\over{E_1}}{\rm e}^{-t_{2}-\sigma_{\rm A}}-{\rm e}^{-t_{\rm A}+\sigma_{\rm A}}.\label{uva2}
\end{align}
Here, $t_{\rm A}$ denotes the time parameter on the anchor curve, and $\sigma_{\rm A}$ is the constant value corresponding to the fixed spacelike coordinate on the anchor curve, as in section 1. In the third equality for $u_{\rm A}$, we use the 2nd-injection time $t_{2}$ defined by 
the relation $v_{\rm A}(t_{\rm 2})={x_{2}/{x_1}}$; see the explanation after \eqref{xxuv}.
The exponential factors ${\rm e}^{2\Phi}$ are evaluated according to \eqref{epi}, $\Phi({\rm A})$ is understood as a function of the coordinates $(u_{\rm A}, v_{\rm A})$. We assume that the boundary point lies in the flat region, so $v_{\rm bdy}<1$, and as a result, ${\rm e}^{\Phi({\rm bdy})}=1$. 

Focusing on the leading contribution, the generalized entropy without the island becomes
\begin{equation}
S_{\rm gen}^\text{no-island}=4E_1\epsilon(t_{\rm A}-\sigma_{\rm A})+\cdots,\label{noisland2injection}
\end{equation}
which increases with respect to $(t_{\rm A}-\sigma_{\rm A})$, as in the single-injection case \eqref{single no island} \cite{Gautason:2020tmk}. This leading contribution does not depend on $E_2$ because the factor $E_1$ originates from the central charge $c$ through the definition $\epsilon\mathrel{:=}c/(48 E_{1})$.

\subsection{With island}
We now consider the case with the island. This is the same as \eqref{1e} with the constant shifts $1\rightarrow 1+E_2/E_1$ and $1\rightarrow 1+(E_2x_1)/(E_1x_2)$. The generalized entropy is given by
\begin{align}
 S_{\rm gen}^{(2,2)}=&2E_1\left\{1+{E_{2}\over{E_1}}-v_{{\rm I}}\left(u_{{\rm I}}+1+{E_{2}\over{E_1}}{x_1\over{x_{2}}}\right)-\epsilon\log(-E_1v_{{\rm I}}u_{{\rm I}})\right\}\notag\\
 +&{c\over{12}}{\log}\left|\left(\log{v_{\rm A}\over{v_{\rm I}}}\log{u_{\rm A}\over{u_{\rm I}}}\right)^2 {\rm e}^{2\Phi({\rm I})}{\rm e}^{2\Phi({\rm A})}
\right|.\label{island2af}
 \end{align}
Here, the superscript $(2,2)$ indicates that both points A and I lie after the 2nd injection. The first term and the second term correspond to the area term and the matter contribution, respectively.  
According to the island formula, $S_{\rm gen}^{(2,2)}$ must be evaluated at the island's saddle point. This requires solving the extremization conditions with respect to $v_{\rm I}$ and $u_{\rm I}$:
 \begin{align}
\partial_{v_{\rm I}}S_{\rm gen}^{(2,2)}=&-2E_1\left(1+{E_{2}\over{E_1}}{x_1\over{x_{2}}}+u_{\rm I}\right)+{2E_1\epsilon\over{v_{\rm I}}}-{8E_1\epsilon\over{v_{\rm I}\log{v_{\rm A}\over{v_{\rm I}}}}}+{4E_1\epsilon\left(1+{E_{2}\over{E_1}}{x_1\over{x_{2}}}+u_{\rm I}\right)\over{1+{E_{2}\over{E_1}}-v_{\rm I}\left(1+{E_{2}\over{E_1}}{x_1\over{x_{2}}}+u_{\rm I}\right)}}=0,\label{vi}\\
\partial_{u_{\rm I}}S_{\rm gen}^{(2,2)}=&-2E_1v_{\rm I}+{2E_1\epsilon\over{u_{\rm I}}}-{8E_1\epsilon\over{u_{\rm I}\log{u_{\rm A}\over{u_{\rm I}}}}}+{4E_1\epsilon v_{\rm I}\over{1+{E_{2}\over{E_1}}-v_{\rm I}\left(1+{E_{2}\over{E_1}}{x_1\over{x_{2}}}+u_{\rm I}\right)}}=0.\label{ui}
 \end{align}
Following the same argument as in \cite{Gautason:2020tmk}, we assume that $\log{v_{\rm A}\over{v_{\rm I}}}$ is sufficiently large. Under this assumption, we can regard \eqref{vi} as a quadratic equation in the variable $v_{\rm I}\left(1+{E_{2}\over{E_1}}{x_1\over{x_{2}}}+u_{\rm I}\right)$. At the leading order, this equation admits two possible solutions: the one is $\epsilon$ and the other is $1+{E_2\over{E_1}}$. The latter solution corresponds to a point near the singularity, so we discard it. The former solution lies near the horizon and will be used in the following analysis.
Using this solution, \eqref{ui} is rewritten as $\log{u_{\rm A}\over{u_{\rm I}}}\sim4\left(1+{E_{2}\over{E_1}}{x_1\over{x_{2}}}+u_{\rm I}\right)$. From \eqref{uva1}, \eqref{uva2} and these relations, the saddle point $(u_{\rm I}(t_{\rm A})\,,v_{\rm I}(t_{\rm A}))$ is written as
\begin{equation}
u_{\rm I}(t_{\rm A}) \sim -1-{E_{2}\over{E_1}}{x_1\over{x_{2}}}-{1\over{3}}{\rm e}^{-t_{\rm A}+\sigma_{\rm A}},\hspace{1cm}
v_{\rm I}(t_{\rm A}) \sim {{3\epsilon}}{\rm e}^{t_{\rm A}-\sigma_{\rm A}}.\label{uvii}
\end{equation} 
Substituting this saddle point into \eqref{island2af}, the generalized entropy with the island becomes
\begin{equation}
S_{\rm gen}^{(2,2)}=2E_1\left\{1+{E_{2}\over{E_1}}-\epsilon(t_{\rm A}-\sigma_{\rm A})\right\}+\cdots, \label{decreentropy}
\end{equation}
which decreases with respect to $(t_{\rm A}-\sigma_{\rm A})$. 
From \eqref{uva1} and \eqref{uvii}, we confirm that the time of the island is delayed compared with the anchor curve, i.e., $v_{\rm I}(t_{\rm A})<v_{\rm A}(t_{\rm A})$. Let us interpret this delay in terms of the time on the anchor curve.
We define the island's time $t_{\rm I}$ by considering the point on the anchor curve satisfying $v_{\rm A}(t_{\rm I})=v_{\rm I}(t_{\rm A})$. 
 Then we obtain the relation $t_{\rm A}=t_{\rm I}+2\sigma_{\rm A}+{\log{1\over{3\epsilon}}}$, which means that $t_{\rm I}$ is delayed compared to $t_{\rm A}$. This delay coincides with the scrambling time \cite{Gautason:2020tmk},\footnote{Since the spacetime structure changes after each ingoing ``null'' energy injection, we consider the island's time in terms of the outgoing null coordinate. Then the island's time $t_{\rm I}$, measured on the anchor curve, is delayed compared with the time $t_{\rm A}$. On the other hand, the scrambling time is the time scale over which a light signal emitted from the anchor curve becomes accessible in the Hawking radiation \cite{Hayden:2007cs, Almheiri:2019psf, Penington:2019npb}. In the island mechanism, it is interpreted as the time scale that the emitted signal hits the island. Under this interpretation, $t_{\rm I}$ which we have introduced corresponds to the time when the signal is emitted from the anchor curve. Thus the difference between $t_{\rm A}$ and $t_{\rm I}$ coincides with the scrambling time \cite{Gautason:2020tmk}. In this setup, both the island's time and the emission time of the signal are related to $t_{\rm A}$ by considering the null coordinate, which is the origin of the coincidence.} defined as the time scale over which the information of the partners of the Hawking quanta appears as the one in the Hawking radiation\cite{Hayden:2007cs, Almheiri:2019psf, Penington:2019npb}.

If the first appearance of the island's endpoint is after the 2nd injection as shown in figure~\ref{pd1}, the Page time is derived by the same argument as the single-injection case.
\eqref{noisland2injection} and \eqref{decreentropy} correspond to \eqref{single no island} and \eqref{singleinjectionisland}, respectively.
Then the Page time is given by $t_{\rm A}-\sigma_{\rm A}={1\over{3\epsilon}}(1+{E_{2}\over{E_1}})$. 
This situation holds in the case where the 1st and the 2nd injections occur simultaneously. This is equivalent to the single-injection case if we define $\epsilon=c/(48(E_1+E_2))$. Although the same situation may hold even in the case with a finite interval between the injections, there could be also different situations in that case.
In the next subsection, we consider the case in which the 2nd injection occurs after the Page time, which has no analogue in the single-injection case. On the other hand, one may consider a case in which the 2nd injection is before the Page time, while the first appearance of the island's endpoint is between the two injections.
A detailed analysis of this case is left for future studies.

\subsection{Intermediate state of the island}\label{jump}
Next, we assume the case in which the 2nd injection occurs after the Page time. Then both the point A at the Page time and the island's endpoint I at its appearance are between the 1st and the 2nd injections. 
Afterwards, the island's delay, explained in the previous subsection, makes the situation possible in which the island's endpoint lies in the region between the two injections, while the point A lies after the 2nd injection, as shown ~\ref{pd2}.
In this subsection, we consider such a situation. If the entropy in this case is favored by the island formula, this state is needed.
In this state, the generalized entropy is given by
\begin{align}
 S_{\rm gen}^{(1,2)}=&2E_1\left\{1-v_{{\rm I}}\left(u_{{\rm I}}+1\right)-\epsilon\log(-E_1v_{{\rm I}}u_{{\rm I}})\right\}\notag\\
 +&{c\over{12}}{\log}\left|\left(\log{v_{\rm A}\over{v_{\rm I}}}\log{u_{\rm A}\over{u_{I}}}\right)^2 {\rm e}^{2\Phi({\rm I})}{\rm e}^{2\Phi({\rm A})}
\right|\label{intsgen},\\
{\rm e}^{2\Phi({\rm A})}=&{v_{\rm A}u_{\rm A}\over{1+{E_{2}\over{E_1}}-v_{\rm A}\left(1+{E_{2}\over{E_1}}{x_{1}\over{x_2}}+u_{\rm A}\right)}}+\order{\epsilon},\notag\\
{\rm e}^{2\Phi({\rm I})}=&{v_{\rm I}u_{\rm I}\over{1-v_{\rm I}\left(1+u_{\rm I}\right)}}+\order{\epsilon}.\notag
 \end{align} 
Here, the only difference from the previous subsection is the part related to $(u_{\rm I},v_{\rm I})$.
From the assumption that the 2nd injection occurs after the Page time,  $t_{2}$ is sufficiently large. This is because, as shown in \cite{Gautason:2020tmk},
the Page time after the first injection is $1/3\epsilon$. Thus the factor ${\rm e}^{-t_{2}}$ becomes sufficiently small.

From the extremization condition, we obtain the relation between $(u_{\rm I},v_{\rm I})$ and $(t_{\rm A}, \sigma_{\rm A})$,
\begin{equation}
u_{\rm I} \sim -1-{1\over{3}}\left({E_{2}\over{E_1}}{x_{1}\over{x_2}}+{\rm e}^{-t_{\rm A}+\sigma_{\rm A}}\right),\hspace{1cm}
v_{\rm I} \sim {3\epsilon}\left({E_{2}\over{E_1}}{x_{1}\over{x_2}}+{\rm e}^{-t_{\rm A}+\sigma_{\rm A}}\right)^{-1}.\label{puvii}
\end{equation} 
Substituting this relation into \eqref{intsgen}, we rewrite
\begin{equation}
S_{\rm gen}^{(1,2)}=2E_1+6E_1\epsilon\log({E_{2}\over{E_1}}{x_{1}\over{x_2}}+{\rm e}^{-t_{\rm A}+\sigma_{\rm A}})+4E_1\epsilon({t_{\rm A}-\sigma_{\rm A}}) +\cdots.
\end{equation} 
Here, the third term is the same as the entropy without the island.
Further taking the derivative with respect to $t_{\rm A}$,
the leading contribution is given by 
\begin{equation}
{d\over{dt_{\rm A}}}S_{\rm gen}^{(1,2)}={2E_1\epsilon\over{1+{E_2\over{E_1}}{\rm e}^{t_{\rm A}-t_2-2\sigma_{\rm A}}} }\left(-1+{2E_2\over{E_1}}{\rm e}^{t_{\rm A}-t_2-2\sigma_{\rm A}}\right)+\cdots.\label{nno}
\end{equation}
Thus $S_{\rm gen}^{(1,2)}$ becomes an extremal value at $t_{\rm A}=t_{2}+2\sigma_{\rm A}+\log{E_1/(2E_{2})}$. From \eqref{nno}, $S_{\rm gen}^{(1,2)}$ is a monotonically decreasing function or a monotonically increasing function in the regime $t_{2}<t_{\rm A}<t_{2}+2\sigma_{\rm A}+\log{E_1/(2E_{2})}$ or $t_{\rm A}\ge t_{2}+2\sigma_{\rm A}+\log{E_1/(2E_{2})}$, respectively.
Here, we assume $2\sigma_{\rm A}+\log{E_1/(2E_{2})}>0$. We discuss the other case in subsection~\ref{nc}.  

The regime $(t_{2}<)t_{\rm A}<t_{2}+2\sigma_{\rm A}+\log{E_1/(2E_{2})}$ implies
\begin{equation}
0<{E_2\over{E_1}}{\rm e}^{t_{\rm A}-t_{2}-2\sigma_{\rm A}}<{1\over{2}}.
\end{equation}
Using this condition, the logarithmic term can be evaluated as
\begin{equation}
\left|\log({E_{2}\over{E_1}}{x_{1}\over{x_2}}+{\rm e}^{-t_{\rm A}+\sigma_{\rm A}})+(t_{\rm A}-\sigma_{\rm A})\right|=\left|\log({E_2\over{E_1}}{\rm e}^{t_{\rm A}-t_{2}-2\sigma_{\rm A}}+1)\right|<\log {3\over{2}}.
\end{equation} 
Thus, the dominant contribution becomes 
\begin{equation}
S_{\rm gen}^{(1,2)}=2E_1\left\{1-\epsilon(t_{\rm A}-\sigma_{\rm A})\right\}+\cdots.
\end{equation} 
Here, this entropy takes the same form as \eqref{singleinjectionisland}.
On the other hand, in the regime $t_{\rm A}> t_{2}+2\sigma_{\rm A}+\log{E_1/(2E_{2})}$, we get the inequality:
\begin{equation}
0<{E_1\over{E_2}}{\rm e}^{-t_{\rm A}+t_{2}+2\sigma_{\rm A}}<2.
\end{equation}
Thus the dominant contribution of $S_{\rm gen}^{(1,2)}$ is given by 
 \begin{equation}
S_{\rm gen}^{(1,2)}=2E_1+4E_1\epsilon(t_{\rm A}-\sigma_{\rm A})-6E_1\epsilon(t_{2}+\sigma_{\rm A})+\cdots.
\end{equation}

The physical meaning of these is interpreted from \eqref{puvii}. 
In the limit $t_{\rm A}\rightarrow \infty$, $v_{\rm I}$ becomes $3\epsilon(E_1/E_2){\rm e}^{t_{2}+\sigma_{\rm A}}$, while $v_{\rm A}$ for the 2nd injection is ${\rm e}^{t_{2}+\sigma_{\rm A}}$.
Here, recall that $E_1$ and $E_2$ are of the same order. 
From this, it follows that $3\epsilon(E_1/E_2){\rm e}^{t_{2}+\sigma_{\rm A}}<{\rm e}^{t_{2}+\sigma_{\rm A}}$. Thus $v_{\rm I}$ doesn't exceed $v_{\rm A}$ for the 2nd injection. This implies that the island's endpoint doesn't reach the 2nd injection, so there can be a time beyond which the island can't catch enough partners of Hawking quanta to decrease the entropy. From the above analysis, such time is indeed found as $t_{\rm A}= t_{2}+2\sigma_{\rm A}+\log{E_1/(2E_{2})}$, and the entropy increases afterwards.

Moreover, there exists a time at which the entropy of the intermediate state becomes equal to the decreasing entropy discussed in the case where both points lie in the region after the 2nd injection \eqref{decreentropy}.
From $S_{\rm gen}^{(2,2)}=S_{\rm gen}^{(1,2)}$,
we obtain the time\footnote{If $\sigma_{\rm A}$ is the variable, \eqref{page intermediate} means that this Page time is null. This is because it can be rewritten as $t_{\rm A}-\sigma_{\rm A}={1\over{ 3\epsilon}}{E_{2}\over{E_1}}+\log{x_2\over{x_1}}$, which is similar to the single-injection case \eqref{tpagesingle}.}  
\begin{equation}
t_{\rm A}={1\over{ 3\epsilon}}{E_{2}\over{ E_1}}+t_2+2\sigma_{\rm A}. \label{page intermediate}
\end{equation}
This time should be regarded as the second Page time, and
at this moment, the entropy is given by
\begin{equation}
S_{\rm gen}^{(2,2)}=S_{\rm gen}^{(1,2)}=2E_1+{4\over{3}}E_2-2E_1\epsilon(t_2+\sigma_{\rm A}).
\end{equation}
Recall that $S_{\rm gen}^{(1,2)}$ increases in the late time, and $S_{\rm gen}^{(2,2)}$ decreases. Thus, before the second Page time, $S_{\rm gen}^{(1,2)}$ is smaller than $S_{\rm gen}^{(2,2)}$, while after the time, $S_{\rm gen}^{(2,2)}$ is smaller than $S_{\rm gen}^{(1,2)}$. Following the island formula, the smallest value must be selected, so the intermediate state is favored before the Page time. Thus the state is needed. 
On the other hand, at the second Page time, the island's endpoint itself moves discontinuously from the region before the 2nd injection to the region after the injection, that is, the island's endpoint jumps, whereas the anchor curve has already been in the region after the 2nd injection.

\subsection{Numerical computation of the entanglement entropy}\label{nc}

\begin{figure}[htbp]
  \centering
  \begin{overpic}[width=0.6\linewidth]{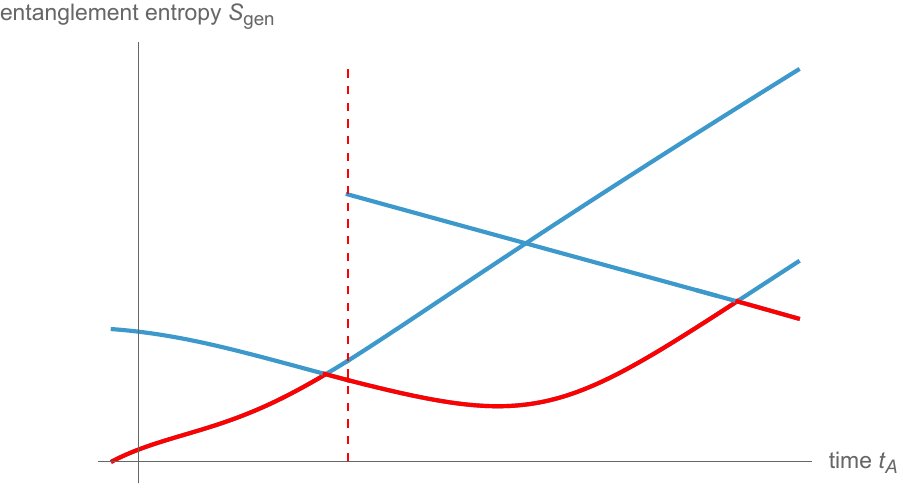}
    \put(30,47){2nd injection}
    \put(43,14){intermediate (1,2)}
        \put(20,18){(1,1)}
          \put(84,14){(2,2)}
    \end{overpic}
  \caption{The Page curve with two injections. The parameters are chosen as follows: $E_1=E_2=3$, $t_2=4.2$, and the other parameters are the same as in figure~\ref{1p}. The red dashed line indicates the time of the 2nd injection. 
The monotonically increasing behavior corresponds to the no-island case. The decreasing behavior before the 2nd injection is the case that both points A and I lie between the injections. For the later argument, we indicate it by writing $(1,1)$. The curve with the local minimum after the 2nd injection corresponds to the intermediate state. The other decreasing behavior after the 2nd injection is the case that both points A and I lie after the 2nd injection. We indicate it by writing $(2,2)$. The red solid line in this graph represents the minimum value, i.e., the Page curve.}
  \label{2p}
\end{figure}
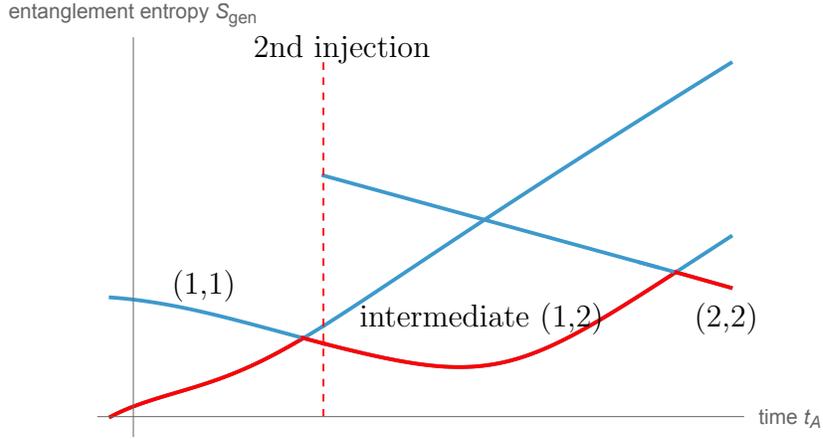

\begin{figure}[htbp]
  \centering
  \begin{overpic}[width=0.6\linewidth]{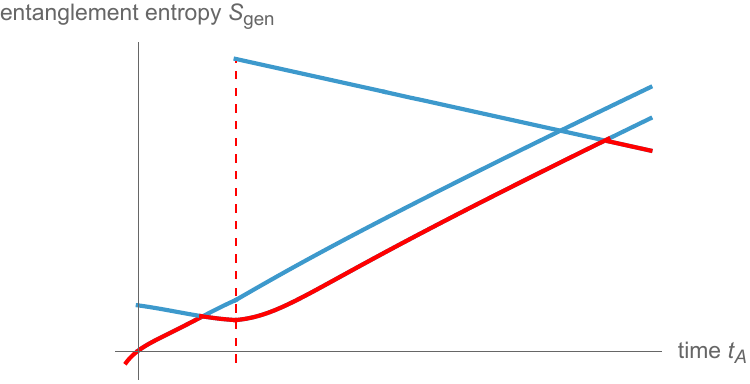}
    \put(25,33){2nd injection}
    \put(58,17){intermediate}
    \end{overpic}
  \caption{The Page curve with two injections using an alternative choice of the parameters: $E_1=3$, $E_2=11.1$, $\sigma_{\rm A}=1$, and the other parameters are the same as in figure~\ref{1p}. Using these parameters, the entropy of the intermediate state monotonically increases.}
  \label{22p}
\end{figure}

\begin{figure}[htbp]
  \centering
   \begin{subfigure}[b]{0.5\linewidth}
     \centering
     \begin{overpic}[width=1\linewidth]{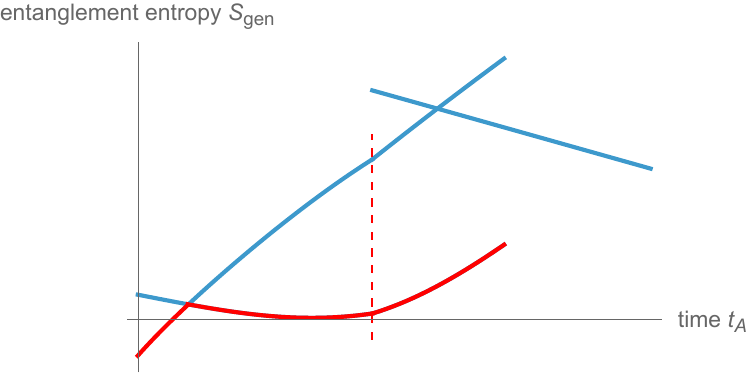}
     \put(35,11){(1,1)}
    \put(60,9){intermediate}
  \end{overpic}
    \caption{}
  \end{subfigure}%
  \begin{subfigure}[b]{0.5\linewidth}
    \centering
     \begin{overpic}[width=1\linewidth]{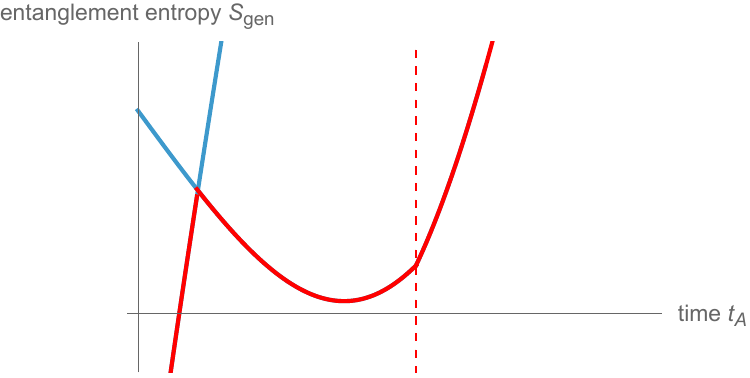}
     \put(38,16){(1,1)}
    \put(64,27){intermediate}
  \end{overpic}
    \caption{}
  \end{subfigure}
  \caption{The Page curve with two injections. We have not succeeded in computing the curve for the late time. The parameters are chosen as follows: $\sigma_{\rm A}= 0.2$, and the other parameters are the same as in figure~\ref{2p}. The label ``(1,1)'' indicates the case that both points A and I lie between the two injections, the corresponding entropy is the curve below (1,1) before the red dashed line. In this case, the entropy with the island exhibits an increasing behavior even before the second injection, unlike in the single-injection case. 
In figure~(a), we understand the entropy of the intermediate state still increases after the 2nd injection,  within the range of our numerical computation. The entropy decreases if both points A and I lie after the 2nd injection. This behavior is the blue decreasing line in figure~(a). We have not confirmed the intersection of the two lines, but we expect the behavior similar to figures~\ref{2p} and \ref{22p}. Figure~(b) highlights the increasing behavior of the entropy with the island even before the 2nd injection.}
  \label{2ppp}
\end{figure}

Next, we numerically compute the entropy. The result is shown in figures~\ref{2p}, \ref{22p} and \ref{2ppp}. To this end, we must consider the five cases:
\begin{enumerate}[label=(\arabic*),itemsep=2pt, topsep=2pt, parsep=0pt, partopsep=0pt]
    \item The no-island case before the 2nd injection.\label{0,1}
    \item Both points A and I lie between the 1st and the 2nd injections.\label{1,1}
    \item The no-island case after the 2nd injection.\label{0,2}
    \item The intermediate state.\label{1,2}
    \item Both points A and I lie after the 2nd injection.\label{2,2}
\end{enumerate}
For each case, we use the entropy given by \eqref{def}, \eqref{1e}, \eqref{ni2i}, \eqref{intsgen}, and \eqref{island2af}, respectively. 
 Although \eqref{def} and \eqref{1e} are used in the single-injection case in section \ref{single}, here we refer to them since they can be used in the case with two injections. 
Note that the anchor curve is given by \eqref{uva1} and \eqref{uva2} in all cases and we take $M=E_1$.
In \ref{0,1} and \ref{0,2}, we describe each generalized entropy as a function of time $t_{\rm A}$.
In all cases with the islands, we perform the following procedure for the numerical computation. First, we analytically derive the extremal conditions like \eqref{vi} and \eqref{ui} for each case.
Then we numerically find the solutions $(u_{\rm I}, v_{\rm I})$ for these conditions and determine the position of the island's endpoint for each appropriately fixed value of $t_{\rm A}$.\footnote{All the numerical computations are done by using the Mathematica, a software system. The Mathematica function ``FindRoot'' is used to find the saddle points. We have also checked that some of the obtained numerical values satisfy both extremal conditions, i.e., 
they correspond to the intersections of the extremal conditions, by 
using the Mathematica function ``ContourPlot''.}
Next we substitute these saddle points into the corresponding generalized entropy, and then vary $t_{\rm A}$ within the appropriate range of $t_{\rm A}$, and obtain each behavior of the entropy.
Finally, we obtain the Page curve by selecting the minimum values among these behaviors, which are plotted as the red lines in figures~\ref{2p}, \ref{22p} and \ref{2ppp}. If it were not for the intermediate state, in other words, if we consider only the existence of both points A and I between two injections or after the 2nd injection, the behavior of the entropy becomes discontinuous at the 2nd injection. Such discontinuity can be seen in figures~\ref{2p}, \ref{22p} and \ref{2ppp} by omitting the intermediate state.%
\footnote{We found this discontinuity to be incorrect owing to the question, at the 2025 Spring meeting of the Physical Society of Japan, about why the entropy is discontinuous.  The author thanks T. Mori for the helpful comment.} By considering the intermediate state, the entropy is continuous in time. 

First,  figure~\ref{2p} corresponds to the case in which $E_1=E_2=3$, $t_2=4.2$, and the other parameters are the same as in figure~\ref{1p}.
In figure~\ref{2p}, as mentioned above, the entropy of the intermediate state decreases before a specific time and increases afterward. After that, the line of the entropy intersects the decreasing one corresponding to the case with both points being after the $2$nd injection. This means that the endpoint of the island moves from the region before the $2$nd injection to the one after the injection.
This intersection corresponds to the Page time, then the entropy changes from increasing behavior into the decreasing one. Note that the island's endpoint does not move across the injection, but it jumps from a point before the 2nd injection to another point after it at the second Page time. See the argument at the end of subsection~\ref{jump}. 

In figure~\ref{2p}, the parameters satisfy the inequality  $t_2<t_{2}+2\sigma_{\rm A}+\log{E_1/(2E_{2})}$. Then the entropy first decreases, and later increases in the intermediate state. 
On the other hand, we may also consider the case in which the time $t_{2}$ of the 2nd injection is after $t_{2}+2\sigma_{\rm A}+\log{E_1/(2E_{2})}$, which corresponds to the condition $2\sigma_{\rm A}+\log{E_1/(2E_{2})}<0$. 
As an example, take $E_2=11.1$, $\sigma_{\rm A}=1$ and the other parameters being the same as in figure~\ref{1p} and figure~\ref{2p}. The result is shown in figure~\ref{22p}.
Then, we observe the only increasing behavior in the intermediate state. Finally, the entropy begins to decrease at the time when the island's endpoint jumps to the region after the 2nd injection.

Moreover, we have observed that there are choices of parameters, with which the entropy $S_{\rm gen}^{(1,1)}$\footnote{Here, $S_{\rm gen}^{(1,1)}$ is the entropy for the case that both points A and I are between the injections.} in the regime before the 2nd injection also changes from the decreasing behavior to the increasing behavior.%
\footnote{The behavior differs from the entropy after the ``first" injection of the single-injection case \cite{Gautason:2020tmk}, in which the entropy only decreases after the Page time. This difference does not mean any discrepancy, since our choice of the anchor curve is different from that of \cite{Gautason:2020tmk}.}
For example, figure~\ref{2ppp}  corresponds to the case $\sigma_{\rm A}=0.2$, while the other parameters are the same as in figures~\ref{1p} and \ref{2p}.
From figure~\ref{2ppp}, we understand the entropy of the intermediate state increases, within the range of our numerical computation.
On the other hand, we observed that the entropy decreases when both points A and I lie after the 2nd injection, again within the range of our numerical computation. This corresponds to the blue decreasing line after the 2nd injection in figure~\ref{2ppp}(a).
Although we have not confirmed the intersection of the two lines, we expect the behavior similar to figures~\ref{2p} and \ref{22p}, i.e., the entropy changes from the increasing behavior to the decreasing behavior after the possible intersection.

In the first two cases we studied, which are depicted in figures~\ref{2p} and \ref{22p}, the entropy repeatedly increases and decreases, and eventually decreases after the second Page time. The way of the repetition depends on the parameters. 
In addition, we expect the similar behavior in the last case which is shown in figure~\ref{2ppp}.

\section{Conclusion}
We discussed the Page curve based on the island formula in the RST model with multiple energy injections.
In section 2, we derived the solution $\Omega$ for the generic $n$, where $n$ is the total number, of the energy injections. From the idea of the apparent horizon based on the dilaton, we confirmed that the horizon expands by the each injection.
In section 3, by using the solution derived in the previous section, we obtained the Page curve based on the island formula, focusing on the case $n=2$.
This extends the analysis of \cite{Gautason:2020tmk} to the two-injection case. Moreover, when the first Page time occurs between the $(n-1)$th and the $n$th injections, the same argument applies to the case with the generic $n$.
Since the spacetime changes with each injection, the entropy analysis using the island formula requires much care. In our analytic calculation, we restricted the point A on the anchor curve to the region after the two injections, but even in this case the position of the island's endpoint could be either after the 2nd injection, or between the two injections. Hence, we need the careful selection of the position of the island's endpoint.
Thus, we considered three cases:
(i) without the island,
(ii) with the island's endpoint between the 1st and the 2nd injections (the intermediate state),
(iii) with the island's endpoint after the 2nd injection.
At the late time in the case (ii), the island can not catch enough partners of Hawking quanta for the entropy to decrease, then the entropy increases.
At the second Page time, the entropy in the case (ii) and the case (iii) become the same value. In the case (ii), the endpoint of the island does not reach the 2nd injection even when the entropy becomes the same value as the case (iii). 
This means that the island's endpoint jumps discontinuously from a point before the 2nd injection to another point after the 2nd injection.
Finally, the entropy starts to decrease after the second Page time. Therefore, the entropy behaves as a continuous function with respect to time. On the other hand, if we forget the intermediate state, the entropy exhibits a discontinuity at the moment of injection. Considering the island's delay, or more precisely the intermediate state, we can derive the continuous Page curve. The intermediate state plays a crucial role for the continuous Page curve.
This is the first analysis to reveal the necessity of the intermediate state and its effect on the continuity of the Page curve under multiple energy injections in asymptotically flat spacetimes. 

In this paper, we clarified aspects of the island. Although the nature of the island itself remains poorly understood,
our study of the island's delay and the intermediate state should contribute to its clarification.
Future work may involve extending this framework to more general spacetimes. We also aim to study the situation where the endpoint of the island lies between the 1st and the 2nd injections while the anchor curve point lies after the 3rd injection.  Additionally, we would like to better understand the island's jump at the Page time.
Ultimately, we aim to deepen our understanding of quantum gravity through the study of the island.

\section*{Acknowledgments}
The author is grateful to A.~Miwa for valuable discussions and continuous guidance. 
The author would also like to thank the members of RIKEN iTHEMS for helpful comments during seminar.
The author thanks the Yukawa Institute for Theoretical Physics at Kyoto University, where this work was completed during the YITP-W-25-07 on ``Strings and Fields 2025''.
Fruitful discussions with participants at the 2025 Spring Meeting of the Physical Society of Japan are also acknowledged.

\end{document}